\documentclass[%
 aip,
 amsmath,amssymb,
 reprint,%
]{revtex4-1}

\usepackage{graphicx}
\usepackage{dcolumn}
\usepackage{bm}

\usepackage[utf8]{inputenc}
\usepackage[T1]{fontenc}
\usepackage{mathptmx}
\usepackage{etoolbox}
\usepackage{lipsum}
\usepackage{xcolor}
\usepackage{gensymb}
 
\makeatletter
\def\@email#1#2{%
 \endgroup
 \patchcmd{\titleblock@produce}
  {\frontmatter@RRAPformat}
  {\frontmatter@RRAPformat{\produce@RRAP{*#1\href{mailto:#2}{#2}}}\frontmatter@RRAPformat}
  {}{}
}%
\makeatother
\begin{document}

\preprint{AIP/123-QED}

\title[Coupling of two individual magnon resonators via a superconducting microwave resonator operating in the strong coupling regime]{Coupling of two individual magnon resonators via a superconducting microwave resonator operating in the strong coupling regime}
\author{Anoop Kamalasanan}
\affiliation{ 
Institut für Physik, Martin Luther Universität Halle-Wittenberg, Halle (Saale), Germany 06120
}%

\author{Georg Schmidt}
\affiliation{ 
Institut für Physik, Martin Luther Universität Halle-Wittenberg, Halle (Saale), Germany 06120
}%
\affiliation{ 
Interdiszinplinäres Zentrum für Materialwissenschaften, Martin Luther Universität Halle-Wittenberg,  Halle (Saale), Germany 06120
}%
\affiliation{
Halle-Berlin-Regensburg Cluster of Excellence CCE, Germany 
}%
 \email{georg.schmidt@physik.uni-halle.de}
\date{\today}
\author{Seth W. Kurfman}%
\affiliation{ 
Institut für Physik, Martin Luther Universität Halle-Wittenberg, Halle (Saale), Germany 06120
}%

\begin{abstract}
We demonstrate coherent coupling of ferromagnetic resonance in two individual permalloy stripes via a superconducting coplanar waveguide resonator. The two stripes are placed on top of the resonator with an angle of 50$\degree$ between their respective long axes. This alignment enables us to separately tune their ferromagnetic resonance frequency by exploiting their shape anisotropy by simply rotating the external bias magnetic field in the sample plane. Both individual ferromagnets exhibit strong coupling with the microwave resonator. This method avoids the use of local individual bias coils and facilitates the integration of similar experiments into planar device infrastructures.

\end{abstract}
\maketitle
Strong coupling of elementary excitations is an important ingredient for quantum information processing and storage. Over the past years, magnons have turned out to be quite attractive in that respect \cite{Lachance-Quirion_2019}, as they can couple to cavity and LC resonator microwave photons \cite{Huebl_2013, Zhang_2014_PRL_StronglyCouplingMagnonsAndCavityMicrowavePhotons, Tabuchi_2014_PRL_HybridizingFerromagneticMagnonsAndMicrowavePhotonsInTheQuantumLimit}, light \cite{Zhu_2020_Optica_WaveguideCavityOptomagnonicsForMicrowaveToopticsConversion, Zhang_2016_PRL_OptomagnonicWhisperingGalleryMicroresonators}, phonons \cite{Zhang_2016_SciAdv_CavityMagnomechanics, An2020, Schlitz2022}, and even to qubits \cite{Tabuchi2015}. The coupling of a single magnon resonator to a single microwave resonator at cryogenic temperatures has, since the preliminary experiments, been demonstrated quite often  \cite{Li2019,Hou2019,Baity2021, Guo_2023_NanoLett_StrongOnChipMicrowavePhotonMagnonCouplingUsingUltralowDampingEpitaxialYIGfilmsAt2K,Xu_2024_AdvScience_StrongPhotonMagnonCouplingUsingALithographicallyDefinedOrganicFerrimagnet, Kurfman2026_arxiv}  as well at room temperature using large microwave cavities \cite{Zhang2016} and split-ring-resonators \cite{Bhoi2014, Zhang2017}. However, the simultaneous control of the coupling of two magnon resonators to (and via) a common microwave resonator has only been demonstrated in a few geometries \cite{Zhang2015, Li2022,Song2025,Pishehvar2025}, where a primary reason for this lack of experiments is their complexity. That is, to tune the high quality magnon resonance to the microwave resonator, an external magnetic field is necessary. To control the coupling of two individual magnon resonators operated at one and the same frequency, however, it is necessary to apply individually tunable magnetic fields to the magnets. The most intuitive way to do this is by using small local bias fields induced by small coils close to the respective magnon hosts.
This kind of experiment has been performed at room temperature with a hollow micro wave cavity \cite{Zhang2015,Pishehvar2025} and also at low temperature with a superconducting resonator which includes two YIG spheres in individual superconducting loops\cite{Li2022,Song2025}. Neither the setups with macroscopic YIG spheres nor the necessity of bias coils, however, make these experiments attractive for integration. 
In order to surpass these limitations, we have developed a simple device geometry in which two permalloy (Py) stripes are arranged with non-collinear long-axes on a superconducting $\lambda/2$ microwave resonator (similar to the one used for example by Li \textit{et al }\cite{Li2019}). This allows us to individually tune the respective ferromagnetic resonance (FMR) frequencies to the microwave resonance and achieve results reminiscent of those obtain in Zhang \textit{et al} \cite{Zhang2015} and Li \textit{et al} \cite{Li2022}.

\begin{figure}[h]
    \centering
    \includegraphics[width=1\linewidth]{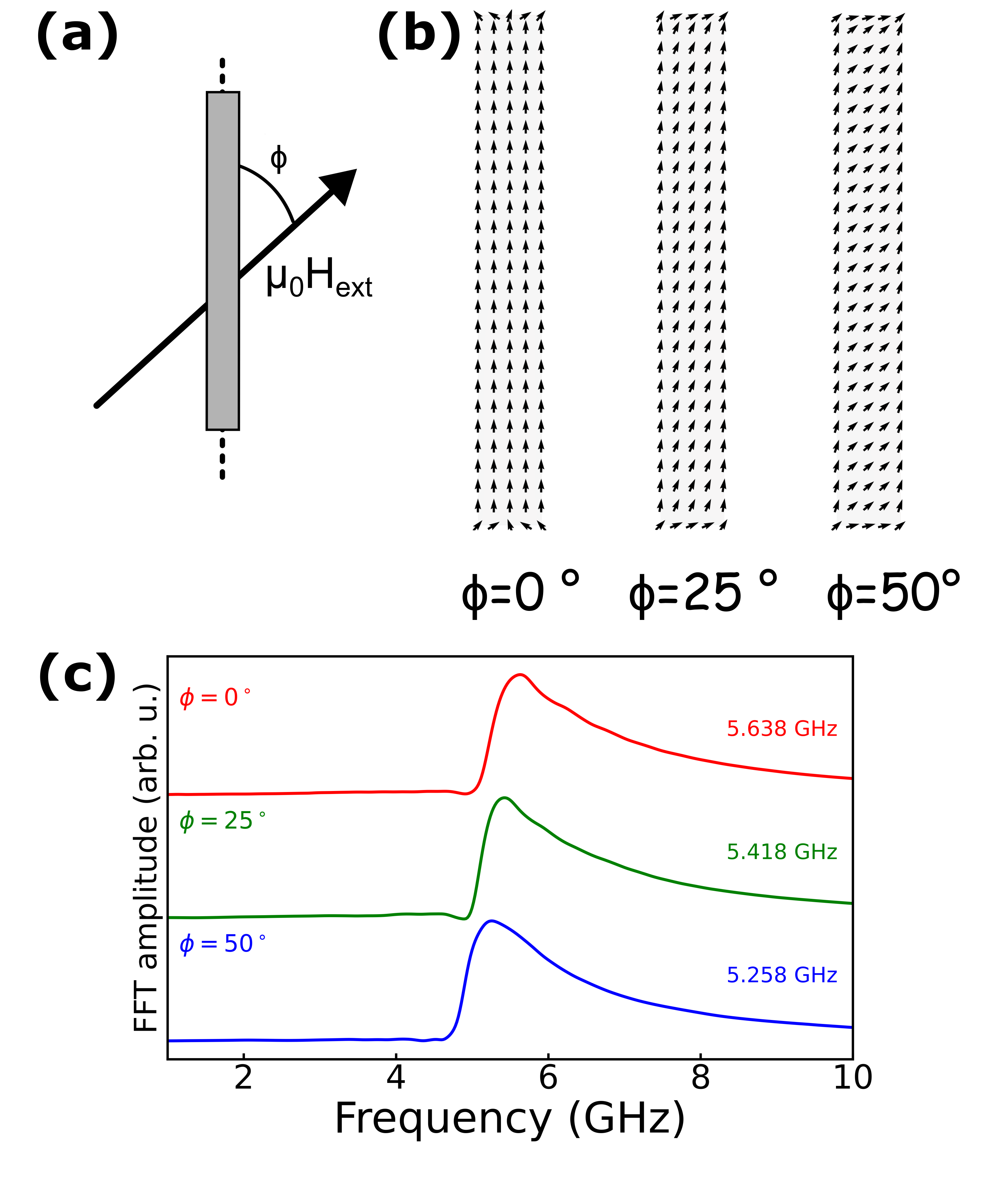}
    \caption{Micromagnetic simulations (mumax$^3$) of the magnetization in a Py stripe for an external magnetic field of $\mu_0 H=30$\,mT under angles of $\phi=0\degree$, $\phi=25\degree$, and $\phi=50\degree$, respectively. (a) shows the geometry defining the angle $\phi$ between the in-plane magnetic field and the long axis of the Py stripe. In (b), the local static magnetization direction in the stripe is shown. One can see that the magnetization follows the external field, however the stripe is not fully saturated. In (c), we see the resonance of the stripe, obtained by a fast Fourier transform of the dynamics in the stripe after excitation by a small, ultrafast magnetic field step. With increasing angle $\phi$, the resonance frequency decreases as a direct consequence of the reduced effective internal magnetic field due to demagnetizing effects. The dimensions of the simulated stripe are $100\,\mu m \times 14\,\mu m \times 30\,nm$. To increase simulation speed, the length of the stripe was reduced with respect to the experiment ($100\,\mu m$ vs. $900\,\mu m$). According to Aharoni \cite{Aharoni1998}, this does not change the basic result because the ratio between length and width of the stripe is larger than 5:1.}
    \label{fig:FIG1_}
\end{figure}

Shape anisotropy is a long-known and well-understood phenomenon \cite{Kittel1948,Aharoni1998} which dictates, for example, that a narrow and thin stripe of ferromagnetic material has an easy axis along its length. Similarly, for such a stripe, a finite magnetic field is necessary in order to magnetize the stripe in a direction which is not parallel to this long axis. The demagnetizing field that causes this anisotropy is part of the effective magnetic field \cite{Kittel1948} $\mu_0 H_{\mathrm{eff}}$ that enters the Landau-Lifshitz-Gilbert (LLG) equation
which determines the precession frequency. Accordingly, using an elongated Py thin-film stripe allows us to exploit its shape anisotropy to vary its dynamic properties in a constant magnetic field simply by rotating the field direction with respect to the long axis of the stripe. The FMR resonance frequency when the field is along the long-axis of the Py is different from when the external field is at an angle to it  (due to the demagnetizing field) and we can thus tune the resonance conditions by a rotation of the field. This is expected and well understood \cite{Kittel1948, Aharoni1998}. We have confirmed the functional phase space by performing micromagnetic simulations  (mumax$^3$)\cite{Vansteenkiste2011,Vansteenkiste_2014_AIP-Adv_TheDesignAndVerificationOfMuMax3} as  outlined in Fig. \ref{fig:FIG1_}. The resonance frequency shifts down (or the necessary resonance field for a set frequency shifts up) when we increase the angle $\phi$ between the magnetic field and the long axis of the stripe. From the simulations, we can also visualize the magnetization profile in the stripe as a function of the applied magnetic field angle and extract the angle of the magnetization with respect to the external field as well as the spatial distribution of the spin dynamics within the stripe. We later utilize this fact by placing two non-collinear permalloy stripes on top of a superconducting resonator (SCR) and deterministically tuning the coupling of the two stripes with the SCR by changing the external field angle. 

\section{\label{sec:level1}Resonator characterization}
For our experiment, we use a superconducting coplanar waveguide (CPW) $\lambda/2$ resonator similar to the ones used by Li \textit{et al} \cite{Li2019} and Hou and Liu \cite{Hou2019}, onto which either one or two Py stripes are placed whose magnetization dynamics can couple to the magnetic microwave field of the resonator. The resonator structure was patterned on a sapphire substrate using photolithography, followed by sputter deposition of a Nb(5 nm)/NbN(200 nm)/Nb(7 nm) multilayer and lift-off. At first, the bare superconducting CPW resonator was characterized via standard vector network analyzer (VNA) transmission spectroscopy ($S_{21}$) techniques. 
\begin{figure*}
    \centering
    \includegraphics[width=1.0\linewidth]{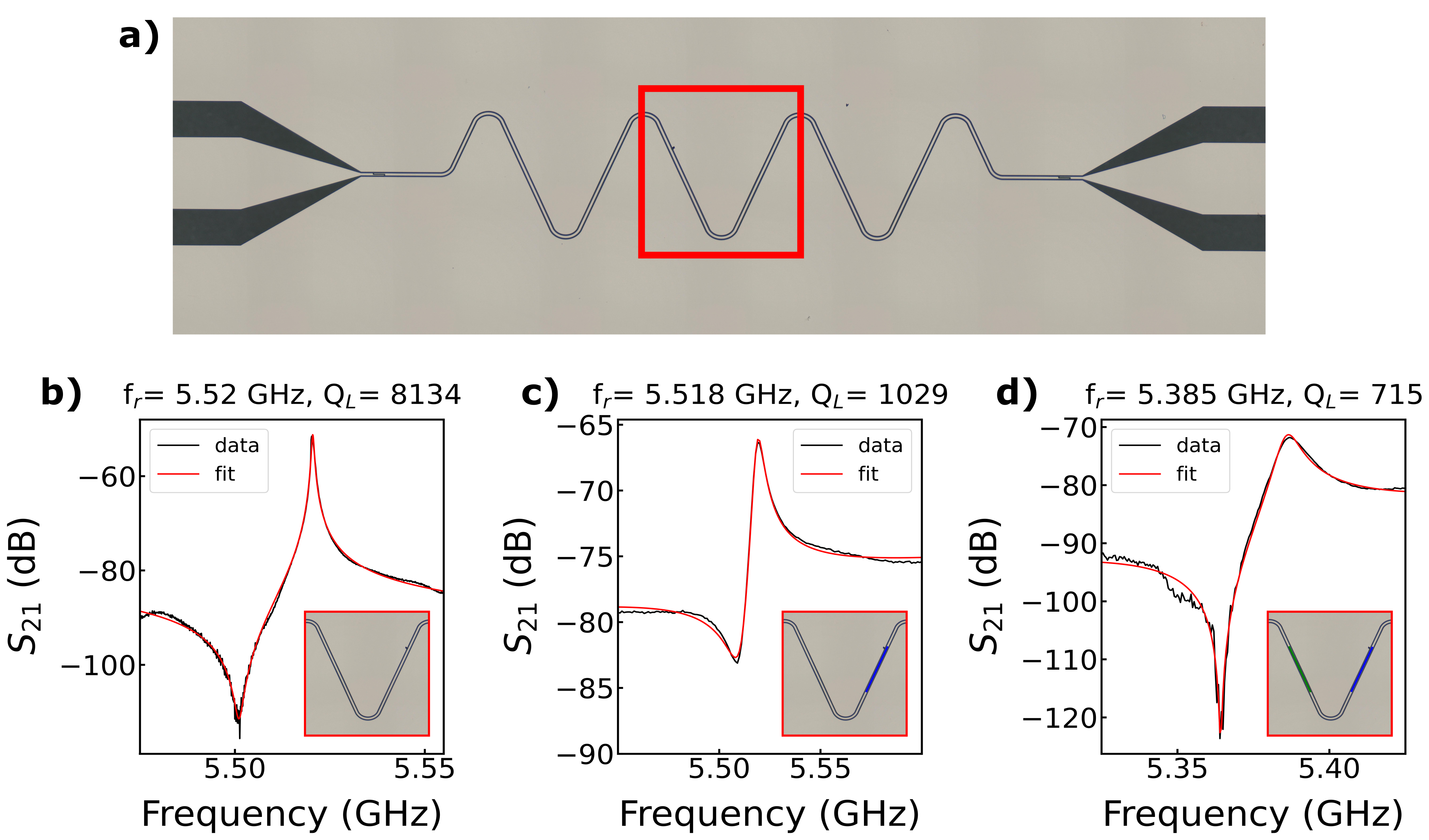}
    \caption{(a) Optical microscope image of the $\lambda/2$ resonator. The red outlined inset shows the region where the Py stripes are patterned. (b-d) $S_{21}$ transmission spectra of the respective resonator sample showing a peak at the resonance frequency of the resonator of (b) the resonator itself, (c) the resonator with a single Py stripe, and (d) the resonator with two Py stripes.}
    \label{fig:FIG2_}
\end{figure*}
The $S_{21}$ microwave response of the resonator was measured at $\approx$1.6 K using a Tektronix TTR506A VNA with an output power of 0 dBm - note that a 20 dB attenuators was placed immediately before and another immediately after the DUT to ensure proper thermal anchoring to the cryogenic environment, and approximately 2.5 dB attenuation of the signal from the cable length. Accordingly, the nominal DUT input power was -22.5 dBm. A resonance of the device, indicated by the peak in the $S_{21}$ signal, was observed at 5.52 GHz with a quality factor $Q \approx$ 8134, indicating relatively low microwave losses (see Fig. \ref{fig:FIG2_}(a)). 

Two more samples were fabricated using the same basic procedure. Following the resonator fabrication, in additional photolithography and sputter-deposition steps, either one or two Permalloy stripes were deposited on top of the respective superconducting resonators, separated from the superconductor by an MgO(20nm) spacer layer to provide electrical isolation. The resonator was designed in order to allow the deposition of two Py stripes with their respective long axis aligned at an angle of  $50\degree$. Microscope images show that the single Py stripe is approx. $12\,\mu$ m x $900 \,\mu$ m x $0.03\,\mu$ m in size. The stripes on the second sample are slightly wider ($13\,\mu m$) but with the same length and thickness. 

For these two samples, the measured quality factors of the resonators were significantly reduced compared to that of the bare resonator. This reduction is caused by the introduction of additional microwave-loss channels associated with the magnetic layer, including magnetic damping, eddy-current losses and possible additional fabrication-induced losses. Figure \ref{fig:FIG2_} shows the respective resonance curves of the devices with asymmetric complex resonator line-shape model fits (red line). As we can see, each Py stripe further reduces the resonance frequency $f_r$ (1 Py: 5.518 GHz , 2 Py: 5.385 GHz) and also the quality factor $Q$ (1 Py: 1029, 2 Py: 715). For the resonator with one stripe, this corresponds to a damping rate of $\kappa/2\pi\approx 5.36$ MHz, while the damping rate for two Py stripes is $\kappa/2\pi\approx 7.53$ MHz.

\begin{figure*}[t]
    \centering
    \includegraphics[width=1\linewidth]{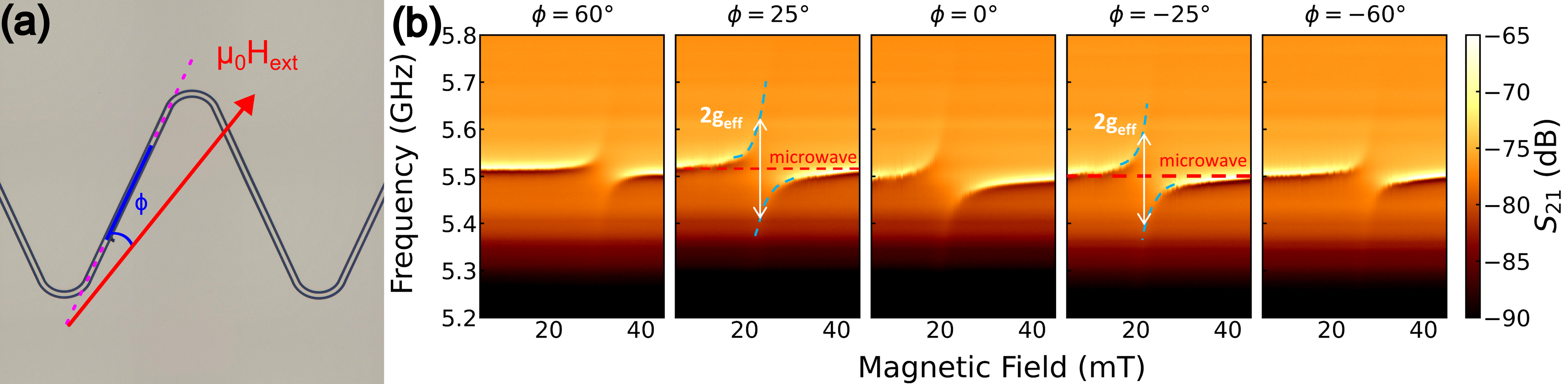}
    \caption{(a) Optical microscope image of the resonator with one Py stripe (indicated in blue) and the magnetic field direction shown by the red arrow.  Stripe and arrow enclose the angle $\phi$. (b) Heat maps of the $S_{21}$ transmission spectra as a function of applied magnetic field and RF-frequency for five different respective magnetic field directions (angles $\phi$). Depending on $\phi$ the position of the avoided crossing changes. The lowest field is observed for $\phi=0\degree$. Note that for $\phi=\pm25\degree$ the resonator frequency and the position of the avoided crossing are ot exacty identical.}
    \label{fig:FIG3_}
\end{figure*}

To characterize the magnon-microwave coupling, frequency sweeps and VNA transmission measurements were done using the same experimental conditions as for the frequency response. By performing these sweeps at different magnetic fields of the device with a single Py stripe (see Fig. \ref{fig:FIG3_}), a two dimensional heat map can be achieved that shows the avoided crossing of the ferromagnetic resonance of the Py and the microwave resonance of the resonator. When the magnetic field is aligned along the permalloy stripe, we expect the minimum resonance field because the demagnetizing field is lowest. At the same time, we expect the strongest coupling because of maximum mode overlap. In this configuration, the gap at the avoided crossing is approx. $280$ MHz yielding a coupling strength of $g_{\mathrm{eff}}/2\pi\approx 140$.
With a magnon damping of $\gamma/2\pi\approx 150$ MHz (FWHM as determined from a control sample)  and a resonator linewidth of $\kappa/2\pi\approx 5.36$ MHz, a splitting of $2g_{\mathrm{eff}}/2\pi\approx280$ MHz gets us well into the strong coupling regime. The corresponding cooperativity can be calculated to\cite{V[TCNE]x-StrCplng,ZareRameshti2022} 
\begin{equation}
C=\frac{4g_{\mathrm{eff}}^2}{\kappa \gamma}\approx 97
\end{equation}
also indicating the regime of strong coupling.

To confirm the viability of our experiment, we now repeat these sweeps for different angles between the magnetic field and the Py stripe. The measurements show that for increasing angles in positive and negative directions the FMR field is increased and the anticrossing moves towards higher magnetic fields (see Fig. \ref{fig:FIG3_}). For an angle of $\pm 90\degree$ (not shown here) the resonance is no longer visible because the RF field is collinear with the external bias field, minimizing the excitation. Without calculating exact numbers the graphs show that as expected the coupling strength is largest for $\phi=0\degree$.

These preliminary measurements on a resonator with a single Py stripe were used to determine a suitable angle between two Py stripes in order to achieve a complete transition between overlapping resonances and complete separation yielding an angle of $50\degree$. This angle between the stripes suggests that in the experiment with two Py stripes, the two stripes have identical resonance frequencies when the field is applied symmetrically under angles of $+25\degree$ to one stripe and $-25\degree$ to the second stripe. Accordingly, we also measure the coupling strength for the resonator with only one Py stripe with the field at angles of $\pm 25\degree$ (see Fig. \ref{fig:FIG3_}(b)). For both angles, we observe the respective avoided crossings as slightly different magnetic fields ($\phi = 25 \degree$ at 24.25mT and for $\phi = -25\degree$ at 21.5mT), and at different coupling strength. While for $\phi=+25\degree$ we have $g_{\mathrm{eff}}/2\pi\approx 145$ MHz we see $g_{\mathrm{eff}}/2\pi\approx 127.25$ MHz for $\phi=-25\degree$.

Both differences are most likely due to the fact that, although $\pm 25\degree$ is symmetric with respect to the Py stripe, it is not with respect to the total resonator. The resonator mainly consists of two sets of differently oriented straight lines connected at the ends in a zigzag pattern. 
While those lines parallel to the permalloy stripes are at $\phi=-25\degree$ or $\phi=+25\degree$ with respect to the magnetic field, the other lines are either at $25\degree$ (for $\phi=-25\degree$) or at $75\degree$ (for $\phi=+25\degree$). The latter means that the magnetic field is almost perpendicular to the lines. The effect on the resonator is visible in Fig. \ref{fig:FIG3_}, where we even observe different respective resonance frequencies for $\phi=-25\degree$ and $\phi=+25\degree$. As a secondary result, one can also expect the quality factor of the resonator to be lower and the geometry of the magnon mode to be different. Both effects may lead to a modification of the coupling strength. 

\begin{figure*}[t]
    \centering
    \includegraphics[width=1\linewidth]{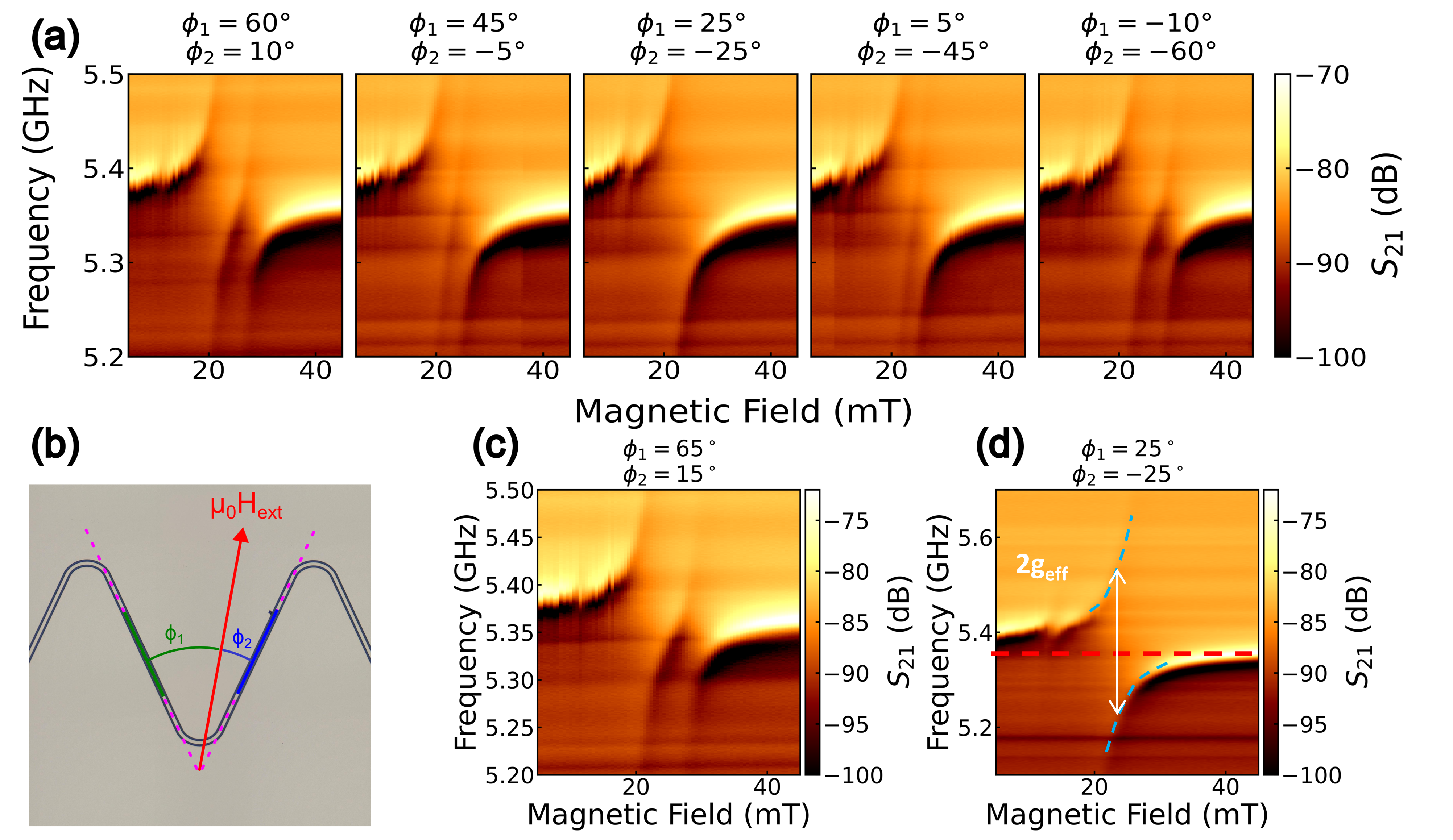}
    \caption{(a) Optical microscope image of the resonator with two Py stripes (one blue and one green) and the magnetic field direction shown by the red arrow. Stripes and arrow enclose the respective angles $\phi_1$ and $\phi_2$. (b) Heat maps of the $S_{21}$ transmission spectra as a function of applied magnetic field and RF-frequency for five different respective magnetic field directions (angles $\phi_1$ and $\phi_2$). For $|\phi_1|=|\phi_2|=25\degree$ only one avoided crossing is visible. When $|\phi_1|$ and $|\phi_2|$ start to differ, a third state in the center of the gap appears. (c) Only for angles of $|\phi_1|=50\degree,55\degree,60\degree,65\degree$ and $|\phi_2|=0\degree,5\degree,10\degree,15\degree$ we observe two separate avoided crossings. (d) Heat map for $|\phi_1|=|\phi_2|=25\degree$ with a larger frequency range. The blue lines are guides to the eye to vizualize the magnitude of $2g_{\mathrm{eff}}$.}
    \label{fig:FIG4_}
\end{figure*}

Subsequently, we characterize the sample with two Py stripes enclosing an angle of $50\degree$. Figure \ref{fig:FIG4_} shows a selection of heat maps taken at different angles $\phi_1$ and $\phi_2$ between the magnetic field and the respective Py stripes. As mentioned before, the fully symmetric case is achieved for $\phi_1=25\degree$ and $\phi_2=-25\degree$. In this case, the respective resonance fields of the two Py stripes are identical and we observe only one avoided crossing. It is noteworthy that the coupling strength (approx. 171.5 MHz) is larger than for a single stripe ($145$ MHz  for $\phi = 25$ and $127.25$ MHz for $\phi = -25\degree$). If we had two Py stripes individually coupled to the resonator but not interacting, the experiment would yield only a higher intensity with the coupling strength for single stripes as we would see a linear superposition of the two results. The fact that the coupling strength is increased indicates that we have an overall coupling for the whole system and that the two Py stripes couple through the microwave resonator. When the angle deviates from this value, the angle of the field with one stripe $\phi_1$ increases while $\phi_2$ decreases for the other stripe. At first this leads to a widening of the gap until a third state seems to appear in the gap center. Only for $\phi_1\gtrapprox50\degree$ we start to see two separate avoided crossings, where each avoided crossing corresponds to the respective resonance of one of the Py stripes. The position of the avoided crossings follows the same systematics as observed for a single Py stripe, thus confirming the initial idea.

Figure \ref{fig:FIG4_} shows the described features in a series of heat maps taken under different angles. The central panel of Fig. \ref{fig:FIG4_}(a) shows the symmetric case ($|\phi_1|=|\phi_2|$) while the outer heat maps were taken for increasing detuning in opposite directions. These results are reminiscent of the demonstration of dark modes for an analog system of a cavity resonator with two YIG spheres as published by Zhang \textit{et al.}~\cite{Zhang2015} Indeed, even when both Py stripes are tuned to the same resonance frequency, a third state must exist in which the resonance dynamics of the two Py stripes are out of phase by $180\degree$. In terms of frequency, this so called dark state is located in the center of the avoided crossing. Because it has no overall dipole moment, it cannot couple to the resonator's magnetic field. In our experiment that measures the power absorption, this state can thus not be excited and remains dark.

In an experiment of coupling between two Py stripes via a superconducting resonator, we have demonstrated that the shape anisotropy of two separate magnetic elements and their alignment with an external magnetic field can be used to tune their FMR frequency though not completely independently, but still in different respective ways. This allows us to tune both stripes to one single or to two different respective resonance frequencies. This method enables us to replace, for example, separate bias coils that are needed to locally adjust individual magnetic field magnitudes. On the one hand this may facilitate experiments on the macro-scale but on the other hand it also allows integration on the microscale. In our experiment, the alignment was achieved using a vector magnet, however, even a simple rotatable sample holder or rotatable magnet can be used to to achieve the tuning. Accordingly, the method is not limited to small 2D samples but can be used with any two magnets that exhibit sufficient shape anisotropy even in macroscopic cavities, although the prospects for integrated coupling experiments are certainly attractive.     

\begin{acknowledgments}
A.K. designed, simulated, fabricated, and measured all devices. A.K., S.K and G.S wrote the manuscript. S.K guided and supported A.K. continuously through initial measurements, simulations, and fabrication. S.K. and G.S. supervised the project, G.S. acquired funding,  and conceived the original project idea.

All authors acknowledge funding by the German Research foundation (DFG) under project number 490952840 (Harmony) and as part of the German Excellence Strategy – EXC3112/1 – 533767171 (Center for Chiral Electronics). 
\end{acknowledgments}

\section*{Data Availability Statement}

The data that support the findings of this study are openly available in Zenodo at http://doi.org/10.5281/zenodo.22125983.

\bibliography{000_2Py_coupling_bibliography}

\end{document}